# Peculiarities of Wigner times delay in slow elastic electron scattering by potential well with arising discrete levels

M. Ya. Amusia[1,2] and A. S. Baltenkov[3]

[1] *Racah Institute of Physics, the Hebrew University, Jerusalem, 91904 Israel*
[2] *Ioffe Physical-Technical Institute, St. Petersburg, 194021 Russia*
[3] *Arifov Institute of Ion-Plasma and Laser Technologies*,
*Tashkent, 100125 Uzbekistan*

**Abstract**
We generalize here the one-level consideration in our recent paper arXiv:1901.00411 [1] to the case when an electron collides with a potential that has any number of s bound states. We investigate peculiarities in the Wigner time delay behavior for slow electron elastic *s*-scattering by spherically symmetric square-potential well. We have considered potential wells, the variation of parameters of which (potential depth *U* and its radius *R*) lead to arising arbitrary number of s bound states. We demonstrate that while the time delay for potential wells with no discrete *s*-levels always has a positive value for small electron energies, it changes sign after level arising. We found that at the moments of arising in the well not only of the first but also following *s*-levels as well, the time delay as a function of *U* experiences instant jumps from a positive value to a negative one. The amplitudes of these jumps increases with decrease of the electron wave vector *k*. The times delay for potential well, the variation of the radius of which *R* leads to the appearance of discrete levels, also change sign at these critical radii.

**1. Introduction**

In a recent paper arXiv:1901.00411 [1] we considered Wigner times delay for slow electron scattering upon a potential, in which variation of parameters leads to formation of loosely bound single *s*- and *p*-states. Here we generalize this consideration, simplifying it at the same time, by investigating the case of a potential that supports arbitrary number of *s* bound states.

Eisenbud, Wigner and Smith (EWS) were the first who interpreted the derivative of the scattering phase shift with respect to particle energy *E* as the time delay of an incident particle wave packet by the scattering potential [2-4]. They introduced the EWS-time delay as a quantum dynamical observable for particle resonance scattering. They have demonstrated that when particle momentum *k* is varied, the capturing and retaining for some time of an incident particle by attractive scattering potential $V(r)$ is replaced by its pushing out, i.e. EWS-time delay is a sign-alternating function of *k*. An important aim of these classical papers was also to obtain maximum information about the phase shift derivatives with a minimum of assumptions concerning a character of the interaction.

To shed some light on special features of the behavior of phase shift derivatives that are completely determined by the shape and parameters of the function $V(r)$ is the aim of our paper. The dependence of scattering phase shifts $\delta_l(k)$ and, for this reason EWS-times delay $\tau_l(k)$, upon the function $V(r)$ can be in latent or explicit form. Let us consider, for the sake of simplicity, electron elastic scattering by a spherical rectangular potential well $V(r)$ with radius *R* and depth *U*. The *s*-scattering phase shift $\delta_0$ for this potential well is described by the following expression (see Eq. (6) in [1])

$$\tan\delta_0 = \frac{k\tan qR - q\tan kR}{q + k\tan qR \tan kR}. \tag{1}$$

The vector $q = \sqrt{2U + k^2}$ here is the electron wave vector inside the potential well[*]. The dependence of the phase shift $\delta_0$ on potential well parameters $R$ and $U$ manifests itself in an explicit form.

The aim of our paper is by using as an example a simple square-potential well $V(r)$ to investigate the connection of the phase shift $\delta_0(k)$ and EWS-time delay $\tau_s(k)$ with the well parameters $R$ and $U$. The reason to consider this problem is as follows. In our paper [5] we studied the partial EWS-times delay in the process of slow elastic electron scattering by the fullerene $C_{60}$ shell. We found there that the sign of partial times delay depends on the presence of the first discrete level with corresponding orbital moment $l$ in the $C_{60}$ potential well. Namely the $l^{th}$ time delay is positive when there are no discrete $l$-levels in the well and it is negative when such a level exists. One can expect similar behavior of the time delay at the moment of appearing in the well not only the first, but also second, third, *etc* discrete levels. To understand whether this specific feature of EWS-time delay behavior is of a universal character, we investigate here the time-delay $\tau_s(k)$ as a function of potential well parameters.

The structure of the paper is as follows. In Section 2 by performing simple calculations we obtain general formulas for the derivative with respect to electron wave vector $k$ of the $s$-phase shift. Section 3 presents results of numerical calculations of the $s$-partial EWS-times delay as functions of $U$ and $R$. Section 4 presents Conclusions.

**2. Main formulas**

Let us rewrite the expression (1) in the equivalent form [6]

$$q \cot qR = k \cot(kR + \delta_0) \tag{2}$$

and apply the operator $\partial/\partial k$ to both sides of this equation. We obtain the following expression

$$q' \cot qR + q(\cot qR)' = \cot(kR + \delta_0) + k[\cot(kR + \delta_0)]', \tag{3}$$

where index prime denotes differentiation with respect to $k$. Bearing in mind that the derivatives in (3) are equal to

$$q' = \frac{k}{q}; \quad (\cot qR)' = -\frac{kR}{q\sin^2 qR}; \quad [\cot(kR+\delta_0)]' = -\frac{(R+\delta_0')}{\sin^2(kR+\delta_0)}, \tag{4}$$

respectively, after performing simple transformations we obtain the following general formula for $\delta_0'$

---

[*] Throughout this paper, we use the atomic system of units (at. un.).

$$\delta_0' = \frac{\partial \delta_0}{\partial k} = -R + \frac{\sin 2(kR+\delta_0)}{2k} + R\frac{\sin^2(kR+\delta_0)}{\sin^2 qR}\left(1 - \frac{\sin 2qR}{2qR}\right). \qquad (5)$$

Let us compare Eq. (5) with the formula (5a) in ref. [3]

$$\delta_0' > -R + \frac{\sin 2(kR+\delta_0)}{2k}, \qquad (6)$$

according to which there exists a general restriction on the behavior of phase derivatives that comes from the principle of causality. The latter states that "the scattered wave cannot leave the scatterer before the incident wave has reached it" [3]. Below we will use Eqs. (5) and (6) in the numerical calculations.

The $s$-partial EWS-time delay is connected with $\delta_0'$ in the following way

$$\tau_s(k,U,R) = 2\frac{d\delta_0}{dE} = 2\frac{d\delta_0}{dk}\frac{dk}{dE} = \frac{2}{k}\delta_0'(k,U,R). \qquad (7)$$

We introduced the well depth $U$ and the radius $R$ as arguments in expression (6) in order to underline that the phase derivative (5) and times delay (6) are functions of three variables, namely $k$, $U$ and $R$. For small electron energy $E = k^2/2$, according to Wigner threshold law [7], the $s$-phase shift is proportional to $k$, $\delta_0(k \to 0) \sim k$. Therefore, $s$-partial EWS-time delay (6) near threshold goes to infinity as $\tau_s(k,U,R)_{k\to 0} \propto \pm 1/k$. Below we will consider functions (5) and (6) for small but finite values of electron energy.

### 3. Numerical calculations
#### 3.1. $U$-dependence

Here we numerically investigate the behavior of the $s$-partial EWS-time delay assuming that the potential well depth $U$ is a variable while the radius $R$ is a constant, equal for definiteness to $R=2$. The condition of appearance of discrete $s$-levels in the potential well is as follows (see Eq. (12) in [4])

$$U^{(n)} = \frac{(2n-1)^2 \pi^2}{8R^2}. \qquad (8)$$

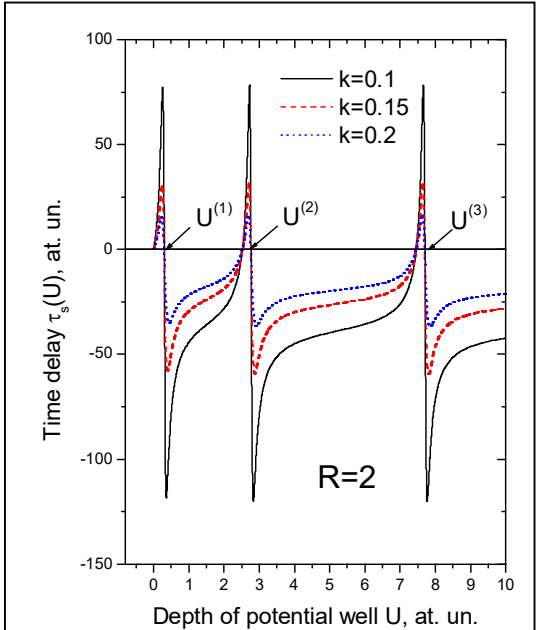

Fig. 1. Time delay $\tau_s(U)$ from Eq. (6) as a function of $U$ for small electron wave numbers $k=0.1$, 0.15 and 0.2. The values of the first critical depths of the potential well with $R=2$ are $U^{(1)} \approx 0.308$; $U^{(2)} \approx 2.776$; $U^{(3)} \approx 7.711$

Here $n$ is the number of the $s$-level in the well. The values of the first critical depths of the potential well with $R=2$ are

$$U^{(1)} = \frac{\pi^2}{32} \approx 0.308; \qquad U^{(2)} = \frac{9\pi^2}{32} \approx 2.776;$$

$$U^{(3)} = \frac{25\pi^2}{32} \approx 7.711. \qquad (9)$$

Figure 1 presents the EWS-time delay $\tau_s(U)$ as a function of $U$ for three small values of the electron wave vectors $k$. For potential wells with no bound levels (the interval $0<U<U^{(1)}$) the time delay has a positive value and changes its sign after the first $s$-level appears in the well with $U$ increase. So, small changes in the potential well parameters (in the given case the well depth $U$) in the vicinity of the critical value $U^{(1)}$ lead to an instant jump of the function $\tau_s(U)$. This specific feature in the behavior of the EWS-time delay is of universal character because we can observe the same picture in the vicinities of other critical values of the variable $U$. The amplitudes of the resonances of curves presented in figure 1 rapidly decrease when the $k$ values increase. When the wave number $k$ increases by a factor of two, from $k=0.1$ to $0.2$, the amplitudes of corresponding curves at the first calculated point after the critical value $U^{(1)}$ differ by more than an order.

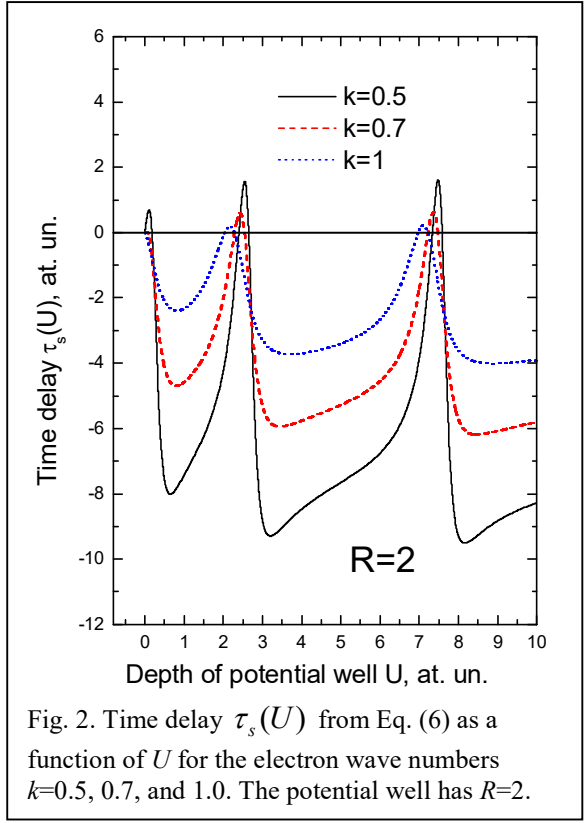

Fig. 2. Time delay $\tau_s(U)$ from Eq. (6) as a function of $U$ for the electron wave numbers $k=0.5, 0.7$, and $1.0$. The potential well has $R=2$.

The relatively simple resonance behavior of the function $\tau_s(U)$ for very small wave number $k$ changes with its further growth that is illustrated by figure 2, that presents results for $\tau_s(U)$ at $k=0.5, 0.7$, and $1.0$. We observe here a much more complex picture. Let us emphasize strongly different scales of oscillations in figures 1 and 2. In the first case the range of the times variation is about $-100$ to $+100$ at. un., while in the second case this range is $-4$ to $+2$ at. un. only. This is an illustration of the following general behavior of the phase shift derivative [8]

$$\delta_0'(k \to \infty) \to 0. \qquad (10)$$

### 3.2. *R*-dependence

In this Section we investigate the behavior of the $s$-partial EWS-time delay assuming that the potential well radius $R$ is a variable while the potential depth $U$ is a constant. The condition of appearance of $s$-discrete levels in the potential well $V(r)$ is given by the following relation (see Eq. (12) in [1])

$$R^{(n)} = \frac{(2n-1)\pi}{2\sqrt{2U}}. \qquad (10)$$

Here $n$, just as above, is the number of the $s$-level in the well. Figure 3 depicts the derivatives of phase shift $\delta_0'(R)$ and EWS-time delay $\tau_s(R)$ as functions of $R$ for fixed electron wave numbers $k$ and potential depths $U$. In the upper panel of this figure the sum $\delta'_0(R) + R$ is presented. According to the general restriction $\delta'_0(R) + R > 0$ (see Eq. (2) in [3]), the curves in this panel should not "penetrate" through the $R$ axis into the lower negative semi-plane, because the scattering center has a finite radius, so that the causality condition can be applied. But according to [3], the wave nature of particles permits some violations of this restriction, and we can see this at the beginning of the curves. These violations disappear with growth of $R$.

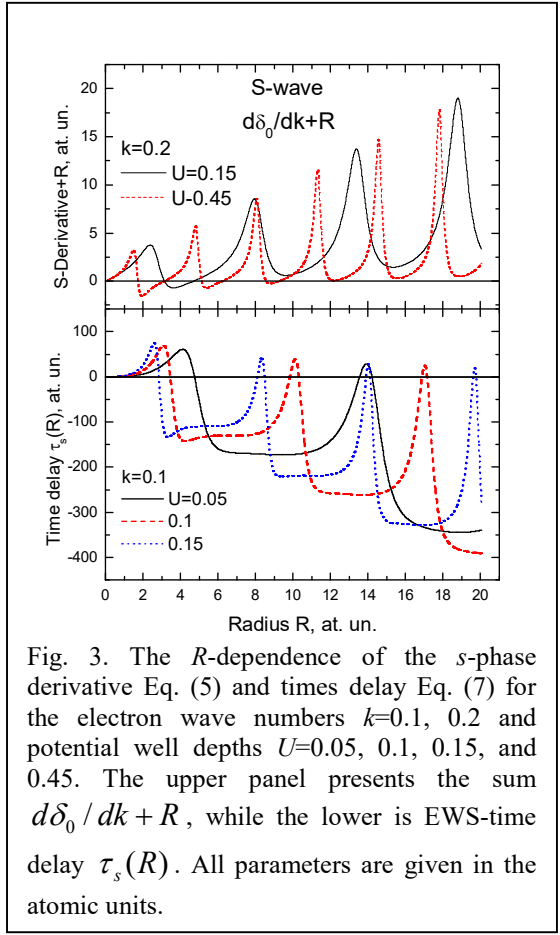

Fig. 3. The $R$-dependence of the $s$-phase derivative Eq. (5) and times delay Eq. (7) for the electron wave numbers $k$=0.1, 0.2 and potential well depths $U$=0.05, 0.1, 0.15, and 0.45. The upper panel presents the sum $d\delta_0/dk + R$, while the lower is EWS-time delay $\tau_s(R)$. All parameters are given in the atomic units.

As in figures 1 and 2, the time delay $\tau_s(R)$ is an alternating function, at least, near the threshold, at small electron energy. However, if in the $U$-dependence case the critical values of potential well depths $U^{(i)}$ coincide with zeroes of curves in figure 1, in figure 3 we see closeness, but not coincidence of zeroes of curves with the critical values of $R^{(i)}$. This is illustrated by Table 1 where the correlation between the positions of the zeroes of the curves on the $R$-axis and the values of $R^{(i)}$ is undoubted.

**Table 1. Comparison between positions of zeroes in curves with $U$ or $R$ variations.**

The critical values of radius $R$ from Eq.(10). The positions of zeroes of the curves in figure 3 are given in brackets. All parameters are in at. un.

| $U$ | $R^{(1)}$ | $R^{(2)}$ | $R^{(3)}$ | $R^{(4)}$ |
|---|---|---|---|---|
| 0.05 | 4.967 (4.72) | 14.902 (14.22) | 24.836 (-) | 34.771 (-) |
| 0.1 | 3.512 (3.42) | 10.537 (10.27) | 17.562 (17.15) | 24.587 (-) |
| 0.15 | 2.868 (2.82) | 8.604 (8.45) | 14.339 (14.10) | 20.075 (19.74) |

### 3.3. $k$-dependence

Figure 4 presents EWS-time delay $\tau_s(k)$ as a function of wave vector $k$. We have two pair of curves. The fist of them, with potential depth $U$=0.15 and radius R=2, corresponds to the case when there is no discrete levels in the well. The second pair corresponds to the potential well with the first s-level ($U$=0.45 at R=2). In addition to the curves calculated with Eq. (5) (solid lines) there are also Wigner functions (WF) calculated with formula (6) (dashed lines). The derivative of the phase shits for slow electron scattering by a potential well with no bound $s$-level is positive and change sign after a first discrete level arises in the well [1]. As in figure 3, the negative parts of dashed curves are manifestations of the scattering particle wave nature. Nevertheless, the restriction $\delta'_0(R) + R > 0$ is essentially preserved also in quantum theory. It does hold, in particular, for large $k$ [3].

Summarizing the results of numerical calculations, one can concluded that the EWS-time delay $\tau_s(k,U,R)$ is an oscillating function relative to all its arguments with a very interesting behavior.

## 4. Conclusions

We have investigated the partial EWS-time delay for slow electron $s$-scattering by rectangular attractive potentials as a function of well parameters depth $U$, radius $R$ and electron momentum $k$. We are concentrated upon vicinity of such parameters of potential wells that are close to their critical values at which $s$-bound states with zero binding energy appear in the wells.

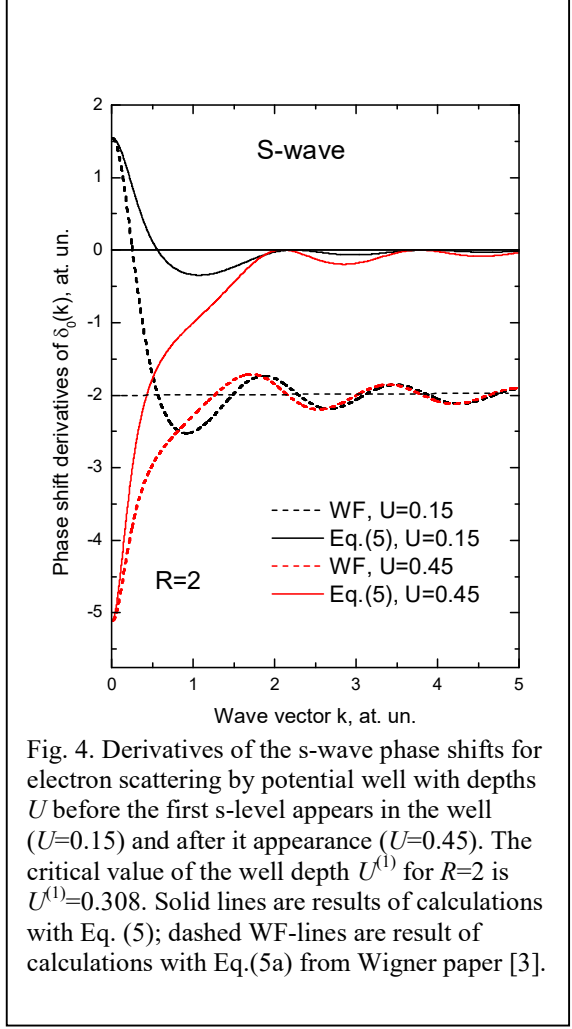

Fig. 4. Derivatives of the s-wave phase shifts for electron scattering by potential well with depths $U$ before the first s-level appears in the well ($U$=0.15) and after it appearance ($U$=0.45). The critical value of the well depth $U^{(1)}$ for $R$=2 is $U^{(1)}$=0.308. Solid lines are results of calculations with Eq. (5); dashed WF-lines are result of calculations with Eq.(5a) from Wigner paper [3].

In spite of its simplicity, the above presented analysis makes it possible to observe some specific features in the time delay behavior that are universal, namely: i) The functions $\tau_s(U)$, $\tau_s(R)$ and $\tau_s(k)$ for slow electron scattering by a shallow potential well with no bound $s$-level is always positive and they change their sign after a first discrete level arises, ii) The small changes of the potential well parameters in the vicinity of the first and any other $s$-levels arising lead to instant jumps of the function $\tau_s(U)$ from a positive value to a negative one. The amplitudes of these jumps increase with decrease of the electron wave number $k$ (see also figure 1 in [1]). This is the specific behavior of $\tau_s(U)$ only; and it is connected with the divergence of the EWS $s$-time near the threshold (see figure 1 in [1]). For orbital moments $l>0$ the EWS-times delay goes to zero near the threshold as $\tau_l(k) \propto k^{2l-1}$ and therefore functions $\tau_l(U)$ smoothly change sign near the critical points on the $U$-axis (see figure 2 in [1]). The function $\tau_s(R)$ change sign as well as $\tau_s(U)$ but this sign alteration, when the radius $R$ is varied, occurs without jumps.

A separate and important problem is to evaluate phase shifts and times not for a simple model but to find time delay in electron-atom scattering processes. A lot is known about scattering phases for this process, for which relatively reliable data on *ab-initio* calculations exist [9], including also low-energy scattering. In a number of cases the addition of attractive polarization interaction alters at small electron energy the $s$-phase derivative from negative Hartree-Fock values to the positive ones. This alteration leads to appearance of the widely known Ramsauer minima in the cross section of slow electron scattering by noble gas atoms. However, it is not connected there to formation of an extra bound electron level. It would be of considerable interest, and this is one of our current directions of activity, to disclose a possible connection between modifications of the atomic field and jumps in $\tau_s(k,U,R)$ on the way from a noble atom to its closest neighbor that is able to form a negative ion.

**Acknowledgments**
ASB is grateful for support to the Uzbek Foundation Award OT-Ф2-46.

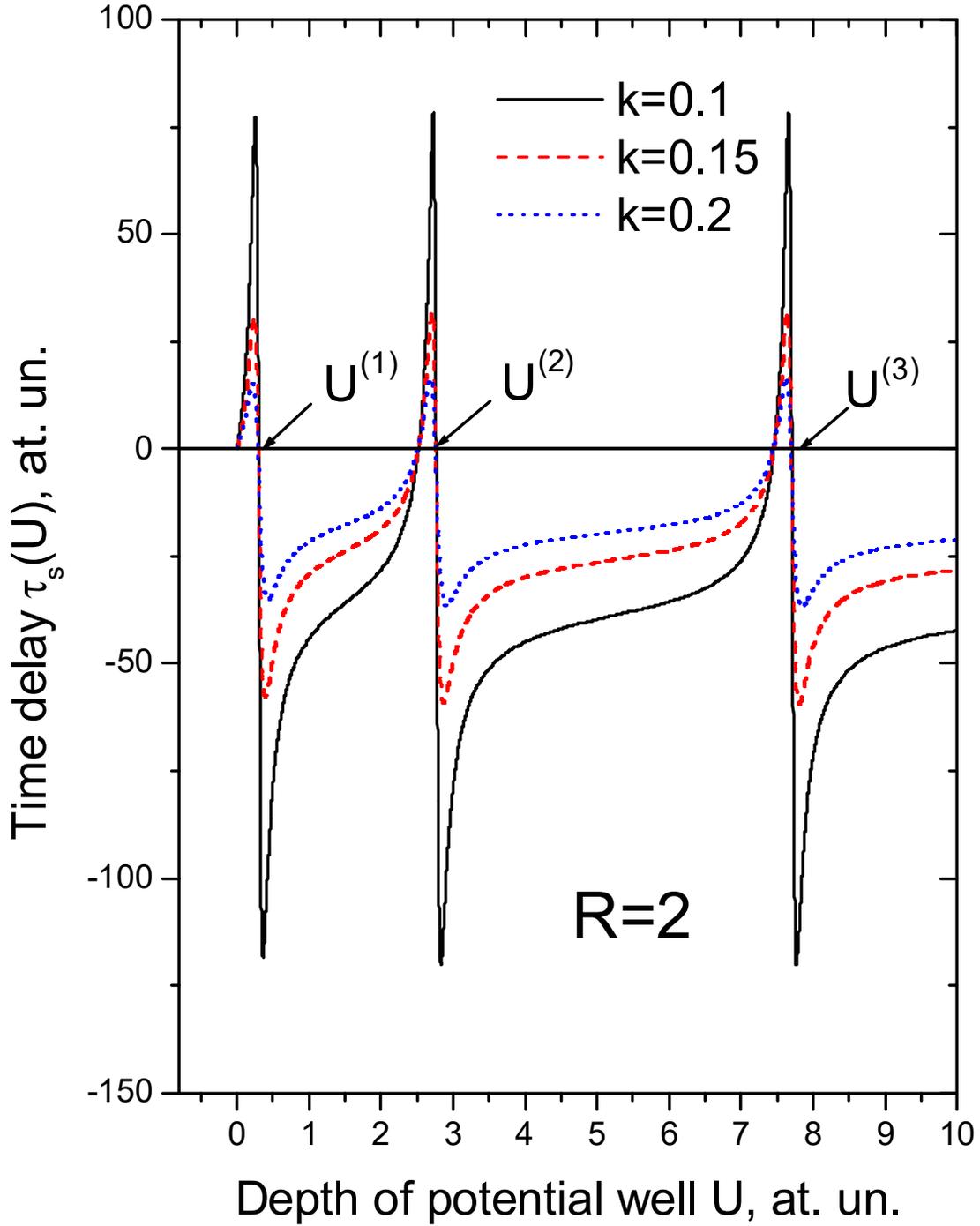

Fig. 1. Time delay $\tau_s(U)$ from Eq. (6) as a function of $U$ for small electron wave numbers $k$=0.1, 0.15 and 0.2. The values of the first critical depths of the potential well with $R$=2 are $U^{(1)} \approx 0.308$; $U^{(2)} \approx 2.776$; $U^{(3)} \approx 7.711$

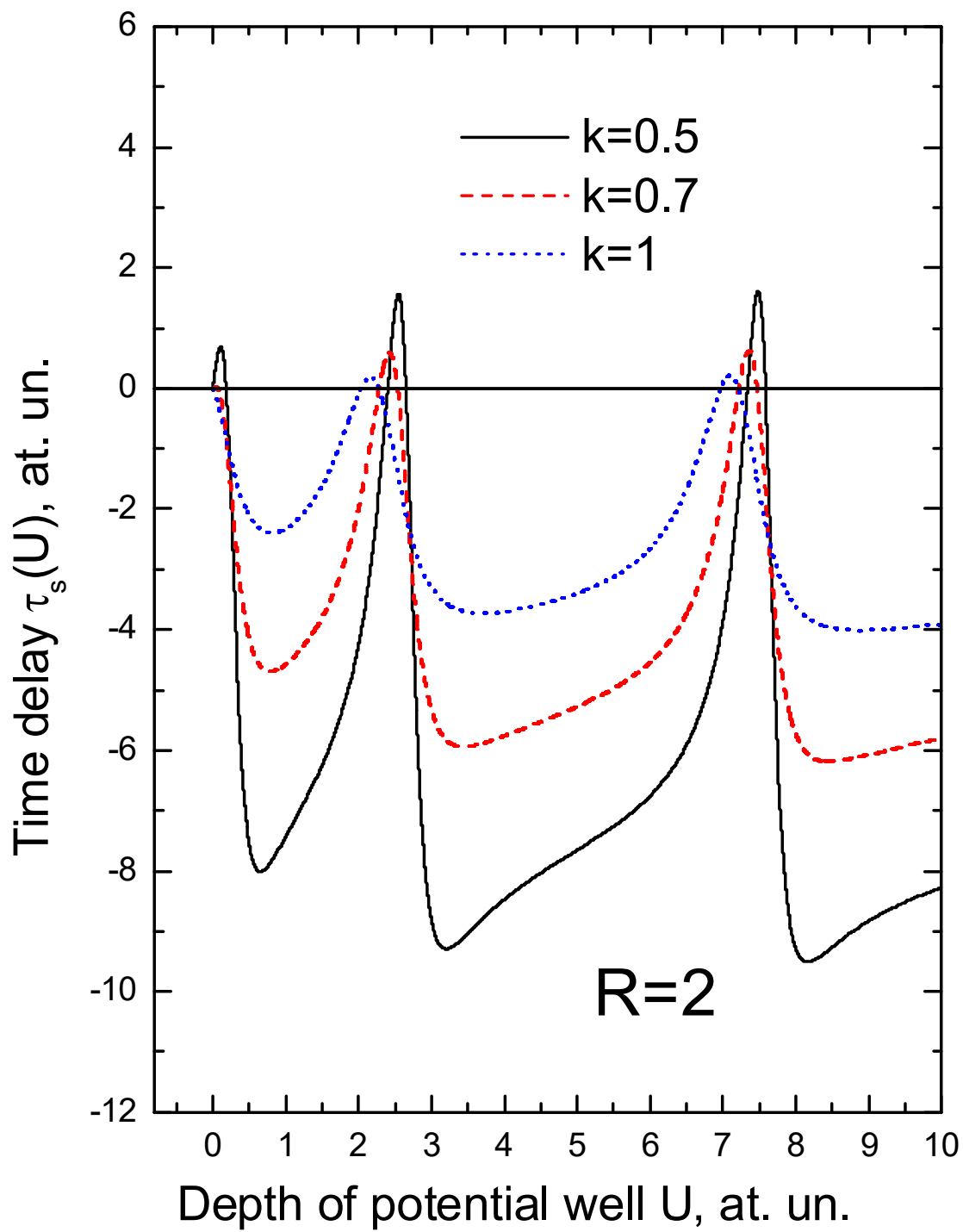

Fig. 2. Time delay $\tau_s(U)$ from Eq. (6) as a function of $U$ for the electron wave numbers $k$=0.5, 0.7 and 1.0. The potential well has $R$=2.

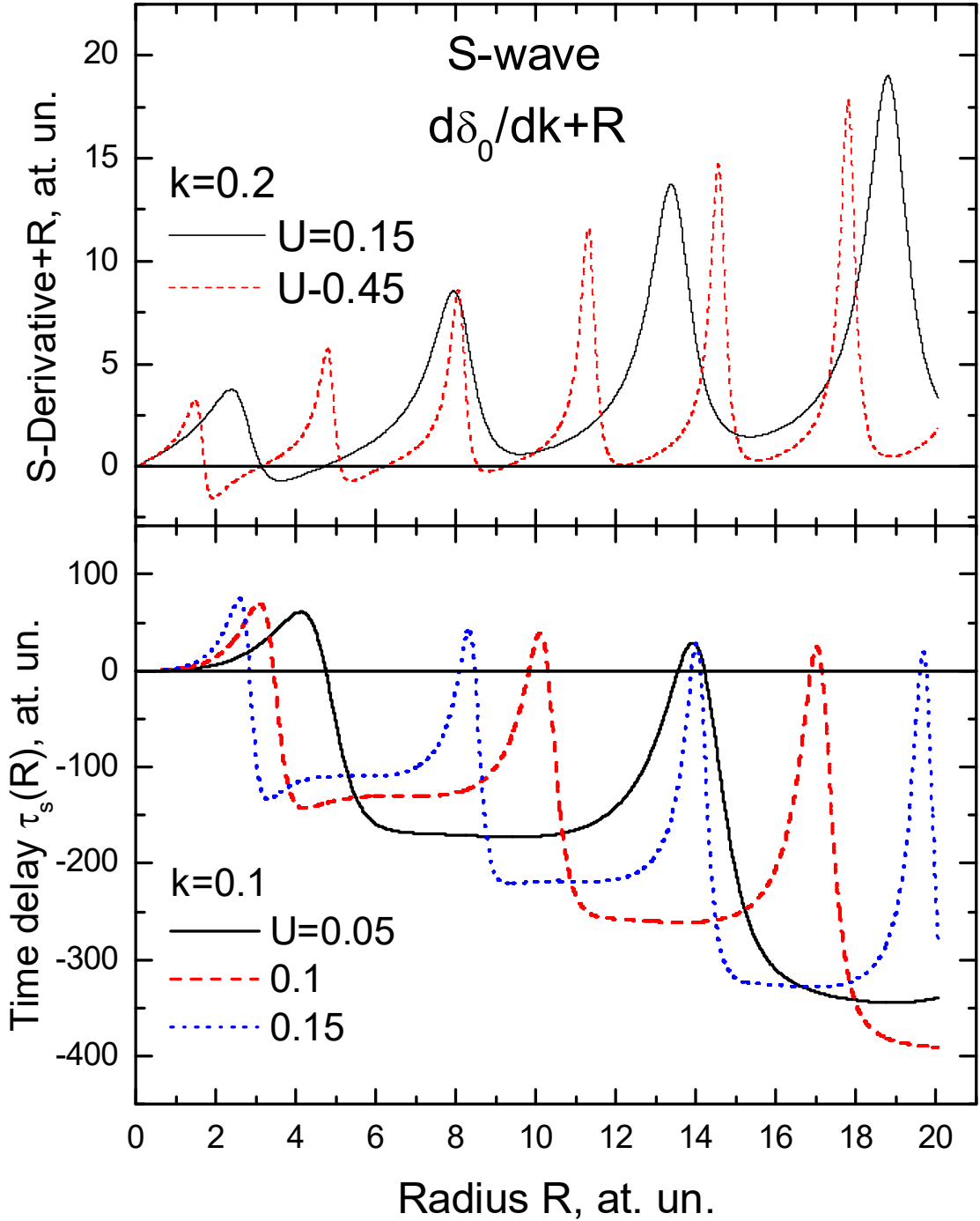

Fig. 3. The *R*-dependence of the *s*-phase derivative Eq. (5) and times delay Eq. (7) for the electron wave numbers *k*=0.1, 0.2 and potential well depths *U*=0.05, 0.1, 0.15, and 0.45. The upper panel presents the sum $d\delta_0/dk + R$, while the lower - EWS-time delay $\tau_s(R)$. All parameters are given in the atomic units.

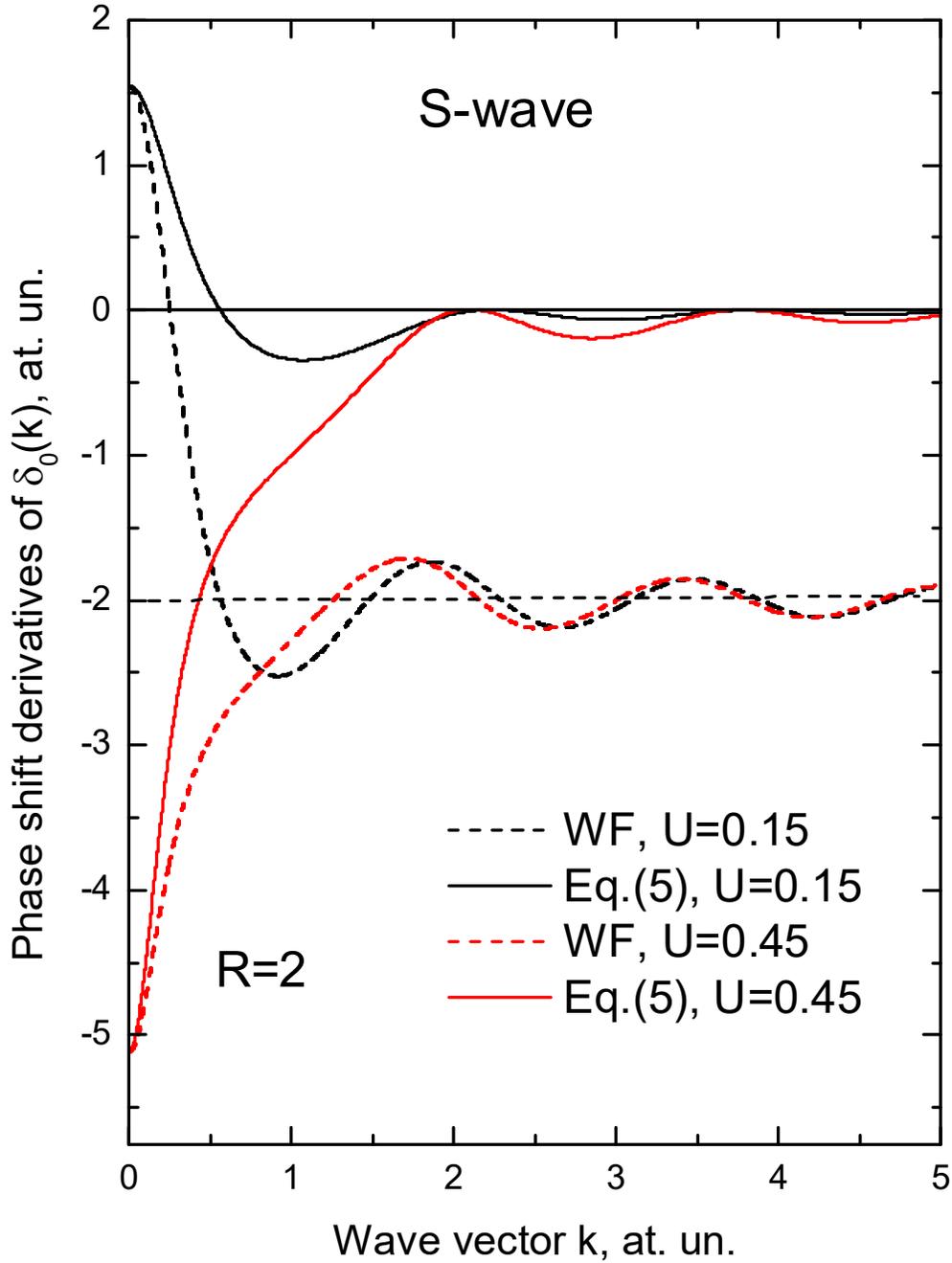

Fig. 4. Derivatives of the s-wave phase shifts for electron scattering by the potential well with depths $U$ before the first s-level appears in the well ($U=0.15$) and after it appearance ($U=0.45$). The critical value of the well depth $U^{(1)}$ for $R=2$ is $U^{(1)}=0.308$. Solid lines are result of calculations with Eq. (5); dashed WF-lines are result of calculations with Eq.(5a) from Wigner paper [3].